\documentclass{aa}

\newcommand{\magpt}[2]{\mbox{$\rm #1\hspace{-0.25em}\stackrel{m}{.}
      \hspace{-1.0mm}#2$}}                             % magnitude \magpt {}{}
\def\bsec{\hbox{$.\!\!{\arcsec}$}}
\def\bmin{\hbox{$.\!\!{\arcmin}$}}
\newcommand\rsec{\hbox{$.\!\!{^{\rm s}}$}}
\newcommand\RA[4]{#1$^{\rm h}$#2$^{\rm m}$#3\rsec#4}

\newcommand\dec[4]{#1$^{\circ}$#2\arcmin#3\bsec#4}
\newcommand\teff{$ {\rm T_{eff}}$}
\newcommand\logM{$\log {\rm M}$}

\newcommand\logg{$\log {\rm g}$}
\newcommand\loghe{${\rm \log{\frac{n_{He}}{n_{H}}}}$}

\newcommand\ebv{$ {\rm E_{B-V}}$}
\newcommand{\Msolar}{\mbox{\,$\rm M_{\odot}$}}        % solar mass
\begin{document}
\thesaurus{05 (08.05.1, 08.08.2, 08.16.3, 10.07.3 NGC~104, 10.07.3 NGC~362)}
\title{Blue horizontal branch stars in metal-rich globular clusters. II. 
47~Tuc and NGC~362}
\author{S. Moehler\inst{1}\thanks{Based on observations collected at the 
European Southern Observatory
(ESO N$^\circ$ 60.E-0145)}
 \and W.B. Landsman\inst{2}
 \and B. Dorman\inst{2}}
\offprints{S. Moehler}
\institute {Dr. Remeis-Sternwarte, Astronomisches Institut der Universit\"at
Erlangen-N\"urnberg, Sternwartstr. 7, 96049 Bamberg, Germany (e-mail:
ai13@sternwarte.uni-erlangen.de)
\and Raytheon ITSS, NASA/GSFC, Greenbelt, MD 20770, USA\\
(e-mail: landsman@mpb.gsfc.nasa.gov, dorman@veris.gsfc.nasa.gov}
\date{Received 3 July 2000 / Accepted 17 August 2000}
\titlerunning{Hot HB stars in 47~Tuc and NGC~362}
\maketitle
\begin{abstract}
Atmospheric parameters (\teff, \logg), and radial velocities are derived
for twelve candidate blue horizontal branch (HB) stars in the
globular clusters 47~Tuc and NGC~362, which so far have been known to
contain primarily red HB stars. The spectroscopic targets were selected
from the catalog of hot stars detected in these clusters at 1600~\AA\ using
the Ultraviolet Imaging Telescope (UIT). Spectroscopic analyses of these
stars reveal, however, that one of the four HB candidates targets in
47~Tuc, and five out of the eight targets in NGC~362 are probably
background stars belonging to the Small Magellanic Cloud. With the 
exception of the photometric binary {\it MJ38529} in 47~Tuc, the parameters of
those stars that are probable members of 47~Tuc and NGC~362 agree well with
canonical HB evolution. The three hot stars in 47 Tuc all have 10,000 K
$<$ \teff $<$ 15,000 K and include one photometric binary, which suggests
that they might have a different physical origin than the dominant red 
HB population. The somewhat cooler blue HB stars in NGC~362 show more
continuity with the dominant red HB population and might simply arise
from red giants with unusually high mass loss. 
\end{abstract}

\section{Introduction\label{47tuc_sec_intro}} 

The colour-magnitude diagrams (CMD's) of globular clusters show a great
variety of horizontal branch (HB) morphologies. It has long been known that
the HB morphology tends to become bluer with decreasing cluster
metallicity. However, the existence of pairs of globular clusters with the
same metallicity but different HB morphologies shows that there must be at
least one ``second parameter''. Theoretically, an increase in age is
expected to reduce the envelope mass and thus yield a bluer HB morphology,
{\em provided that no other parameters are changing.} If age {\em is} the
second parameter, then there are profound consequences for models of the
formation of the galactic halo (e.g.\ Sarajedini et al. \cite{sara95}) and
the origin of the ultraviolet upturn in elliptical galaxies (e.g.\ Park \&
Lee \cite{park97}). However, much recent work indicates that age may not be
the only or even the dominant second parameter controlling the HB morphology
(Stetson et al. \cite{stet96}, Ferraro et al. \cite{ferr97}, 
Catelan \cite{cate00}). Fusi Pecci et
al. (\cite{fusi93}) suggested that the cluster density may influence the
extension of the blue HB to rather faint magnitudes (``blue tail'') that is
seen in the CMD's of many globular clusters, due to enhanced mass loss
triggered by stellar interactions (see also Bailyn \cite{bail95}). 

The clusters 47 Tuc and NGC 362 are famous for their red HB morphologies.
47 Tuc is the prototype metal-rich ([Fe/H] = $-$0.71; Harris
\cite{harr96}) red HB cluster, with only a single known RR Lyr star (V9;
Carney et al. \cite{cast93}). NGC 362 [Fe/H = $-$1.16;  Harris
\cite{harr96}) is the red member of a famous second-parameter pair of
clusters (along with NGC 288). Thus it was a surprise when observations
with the Ultraviolet Imaging Telescope (UIT) revealed the presence of 51
hot (\teff\ $>$ 8500~K) stars in the field of 47 Tuc (O'Connell et al.
\cite{ocdo97}) and 84 hot stars within a radius of 8\bmin25 from the center
of NGC 362 (Dorman et al. \cite{dosh97}). Although both 47~Tuc and NGC~362
are located in projection on the outer halo of the Small Magellanic Cloud
(SMC), O'Connell et al. and Dorman et al. were able to show from the
distribution with cluster radius that a significant fraction of the
detected hot stars should be cluster members. Independently, Kaluzny
et al. (\cite{kakr97}, \cite{kaku98}) used optical photometry to identify
eight candidate blue HB stars in 47~Tuc (all of which were detected with
UIT), but suggested from the absence of a radial gradient that most of
these candidates were actually SMC halo members. 

If the hot stars in NGC 362 and 47~Tuc are HB stars and cluster members,
then they should have the same age and metallicity as the majority red HB
population, and might provide important clues to the other
parameters that can yield blue HB stars in globular clusters. However, as
noted above, not all the hot stars detected on the UIT images are cluster
HB stars. From the distribution of young stars in the outer halo of the SMC
mapped by Gardiner \& Hatzidimitriou (\cite{gaha92}), we expect significant
SMC contamination of the hot stars in NGC 362, whereas 47 Tuc is located in
a region mostly devoid of young stars in the northwest SMC halo. Some of
the hot stars in NGC 362 and 47 Tuc might also be extreme blue stragglers,
analogous to those seen in the cores of NGC 6397 (Burgarella et al.
\cite{bupa94}) and M3 (Ferraro et al. \cite{fepa97}). The temperatures of
these blue stragglers can reach 12,000~K, and recent HST spectroscopy has
confirmed that the mass of the two bright blue stragglers in the core of
NGC 6397 exceeds twice the turnoff mass (Sepinsky et al. \cite{sesa00}) . 

To reach any conclusions about these stars and their evolutionary status we
need information about their atmospheric parameters (effective temperature
and surface gravity) and also about their cluster membership. We therefore
started a programme to obtain spectra of the hot HB star candidates in 47
Tuc and NGC 362. The radial velocities can be used to assess cluster
membership, and the derived atmospheric parameters (\teff\ and \logg) can
be used to distinguish HB stars from blue stragglers and background stars. 

We describe the observations and their reduction Sects.~\ref{47tuc_sec_obs}
and \ref{47tuc_sec_redu}. The determination of atmospheric parameters is
described in Sect.~\ref{47tuc_sec_par} and the results are discussed in
Sect.~\ref{47tuc_sec_disc}. 

\section{Observations\label{47tuc_sec_obs}}

\begin{figure}
\vspace{10.2cm}
\includegraphics{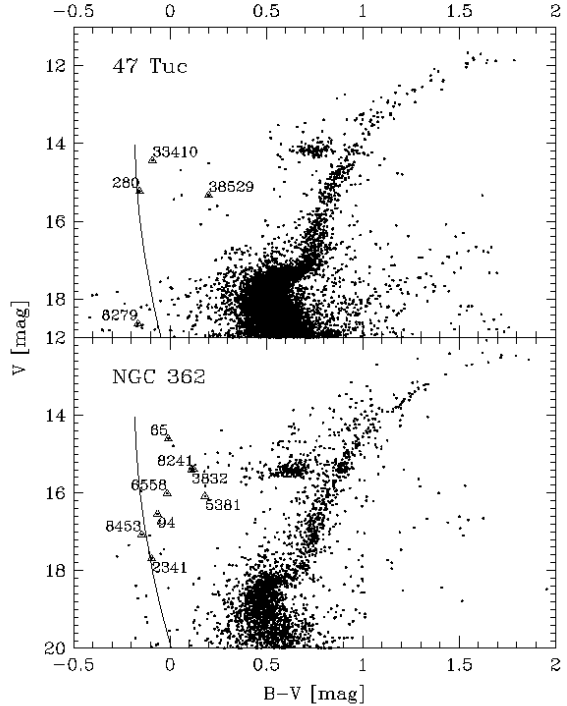}
\caption{The colour-magnitude diagrams for 47~Tuc and NGC~362 as observed
by Montgomery \& Janes (priv. comm.). The triangles mark the stars
for which we took spectra (with the numbers referring to
Table~\ref{47tuc_n362_targ}). The solid line marks the main sequence of
the Small Magellanic Cloud, as derived from the interior models of
Charbonnel et al. (\cite{chme93}) using the colour transformation of Kurucz
(\cite{kuru92}) and $(m-M)_0$ = \magpt{18}{91} and \ebv\ = \magpt{0}{09}
for the SMC. \label{cmd_47tuc_n362}} 
\end{figure} 

We selected a total of 12 blue stars in 47~Tuc and NGC~362 that were
sufficiently uncrowded to allow ground-based follow-up spectroscopy. These
stars are listed in Table~\ref{47tuc_n362_targ} and marked in
Fig.~\ref{cmd_47tuc_n362} with triangles. 

\begin{table*}
\caption[]{Coordinates, photometric data, distance from cluster center, 
heliocentric radial velocities
(obtained from our spectra), and cluster membership information (from 
proper motions) for the
target stars in 47~Tuc and NGC~362. The numbers and the $B, B-V$
photometric data are from Montgomery \& Janes (priv. comm.). $m_{162}$ is
the UIT magnitude at 1620~\AA . The proper motion
cluster membership is taken from
Tucholke (\cite{tuch92a}, \cite{tuch92b}).\label{47tuc_n362_targ}} 
\begin{tabular}{r|ll|rrrr|rrc}
\hline
Star & $\alpha_{2000}$ & $\delta_{2000}$ & $B$  & $B-V$ & 
$m_{162}$ & $m_{162}-V$ & dis & v$_{\rm rad, hel}$ & proper motion \\ 
 & &  & & & & & [$'$]& [km s$^{-1}$] & member (probability)\\
\hline
\multicolumn{10}{c}{47 Tuc}\\
\hline
{\it MJ280} & \RA{00}{21}{29}{35} & \dec{$-$71}{58}{14}{3} & 
\magpt{15}{07} & \magpt{-0}{16} & \magpt{13}{18} & \magpt{-2}{05} 
   & 5.6 & $-$37 & 96.8\% (T1948)\\
{\it MJ8279} & \RA{00}{22}{57}{71} & \dec{$-$71}{57}{56}{6} & 
\magpt{18}{47} & \magpt{-0}{17} & \magpt{15}{86} & \magpt{-2}{78} 
   & 13.7 & $+$161 & ---\\
{\it MJ33410} & \RA{00}{24}{44}{79} & \dec{$-$72}{09}{33}{1} & 
\magpt{14}{34} & \magpt{-0}{09} & \magpt{13}{70} & \magpt{-0}{73} 
   & 9.0 & $-$25 & ---\\
{\it MJ38529} & \RA{00}{25}{33}{20} & \dec{$-$72}{10}{44}{3} & 
\magpt{15}{54} & \magpt{+0}{20} & \magpt{13}{82} & \magpt{-1}{51} 
   & 8.6 & $-$46 & 97.8\% (T300)\\
\hline
\multicolumn{10}{c}{NGC 362}\\
\hline
{\it MJ65} & \RA{01}{01}{58}{4 } & \dec{$-$70}{50}{28}{9} & 
\magpt{14}{59} & \magpt{-0}{01} & \magpt{14}{97} & \magpt{+0}{37} 
  & 6.2 & $+$120 & 18.3\% (T15)\\
{\it MJ94} & \RA{01}{03}{18}{7 } & \dec{$-$70}{52}{20}{9} & 
\magpt{16}{47} & \magpt{-0}{07} & \magpt{15}{04} & \magpt{-1}{50} 
  & 6.1  & $+$137 & 0.0\% (T16)\\
{\it MJ2341} & \RA{01}{03}{11}{8 } & \dec{$-$70}{51}{20}{8} & 
\magpt{17}{59} & \magpt{-0}{10} & \magpt{15}{29} & \magpt{-2}{40} 
  & 3.9  &  $+$75 & --- \\
{\it MJ3832} & \RA{01}{03}{04}{5 } & \dec{$-$70}{51}{30}{5} & 
\magpt{15}{51} & \magpt{+0}{11} & \magpt{15}{19} & \magpt{-0}{21} 
  & 1.0 & $+$197 & --- \\
{\it MJ5381} & \RA{01}{04}{17}{8 } & \dec{$-$70}{54}{23}{8} & 
\magpt{16}{26} & \magpt{+0}{18} & \magpt{16}{38} & \magpt{+0}{30} 
  & 5.5 & $+$109 & 0.2\% (T247)\\
{\it MJ6558} & \RA{01}{03}{16}{1 } & \dec{$-$70}{51}{27}{1} & 
\magpt{16}{00} & \magpt{-0}{02} & \magpt{14}{66} & \magpt{-0}{36} 
  & 1.6 & $+$228 & 0.0\% (T304)\\
{\it MJ8241} & \RA{01}{03}{25}{0 } & \dec{$-$70}{51}{55}{2} & 
\magpt{15}{51} & \magpt{+0}{12} & \magpt{15}{74} & \magpt{+0}{35} 
   & 2.4 & $+$244 & ---\\
{\it MJ8453} & \RA{01}{02}{00}{1 } & \dec{$-$70}{51}{15}{0} & 
\magpt{16}{92} & \magpt{-0}{15} & \magpt{15}{90} & \magpt{-2}{17} 
   & 4.0 & $+$73 & 0.0\% (T396)\\
\hline
\end{tabular}

\begin{tabular}{l}
v$_{\rm rad, SMC}$ = $+$149 km s$^{-1}$ (Hatzidimitriou et al. \cite{hacr97})\\
v$_{\rm rad, 47 Tuc}$ = $-$19 km s$^{-1}$ (Pryor \& Meylan \cite{prme93})\\
v$_{\rm rad, NGC 362}$ = $+$223 km s$^{-1}$ (Pryor \& Meylan \cite{prme93})\\
\end{tabular}
\end{table*}

We observed with the {\it ESO Multi Mode Instrument} (EMMI) at the {\it New 
Technology Telescope} (NTT) on October 26 and 27, 1997, using only the
blue channel as the use of the two-channel mode leads to destructive
interference around the H$_\beta$ line. We obtained low resolution
spectrophotometric data with large slit widths to analyse the flux
distribution and medium resolution spectra to measure Balmer line profiles
and helium line equivalent widths. Seeing values for these observations
varied between 0\bsec7 and 1\bsec3, but the nights were not photometric.
Since no EMMI grism allows observations below 3600~\AA\ (necessary to
measure the Balmer jump) we used grating \#4 (72 \AA/mm) and reduced the
dispersion for the low resolution spectra by binning along the dispersion
axis by a factor of 4. We used a slit width of 1\bsec0 (5\bsec0) for the
medium (low) resolution spectra and the slit was kept at parallactic angle
for all observations. The CCD was a Tek 1024$\times$1024 chip with
(24$\mu$m)$^2$ pixels, a read-out-noise of 5.7e$^-$ and a gain of 2.84
e$^-$/count. 

For calibration purposes we observed each night 10 bias frames and 10 dome
flat-fields with a mean exposure level of about 10,000 counts each. In
addition we obtained sky flat fields to correct for the slit illumination.
We took wavelength calibration spectra before and after each science
spectrum. As flux standard stars we used LTT~7987 and EG~21. 

\begin{figure}
\vspace{10.cm}
\includegraphics{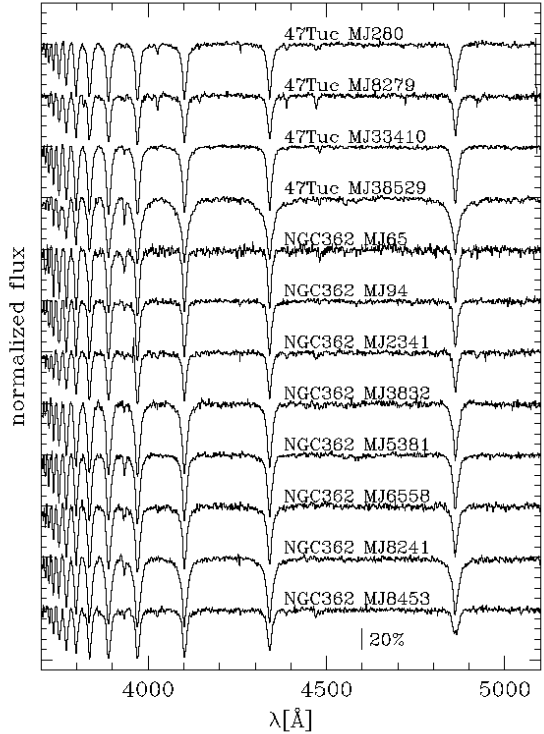}
\caption{The normalized medium resolution spectra of the target stars in 
47~Tuc and NGC~362. The part shortward of 3900~\AA\ was normalized by 
taking the highest flux point as continuum value.\label{47tuc_n362_spectra}}
\end{figure}

\section{Data Reduction\label{47tuc_sec_redu}}

We averaged the bias frames of the two nights and used their mean value
instead of the whole frames as there was no spatial or temporal variation
detectable. To correct the electronic offset we adjusted the mean bias by
the difference between the mean overscan value of the science frame and
that of the bias frame. The dark currents were determined from several long
dark frames and turned out to be negligible (3$\pm$2 counts/hr/pixel). 

The flat fields were averaged separately for each night, since there was a
slight variation in the fringe patterns of the flat fields from one night
to the next (below 5\%). We averaged the sky flat fields and then condensed
them along the dispersion axis to derive the spatial profile of the slit.
The dome flat fields were treated the same way, divided by their spatial
profile and then multiplied by the spatial profile derived from the sky
flat fields. The spectral energy distribution of the flat field lamp was
determined by averaging the dome flat fields along the spatial axis. This
one-dimensional ``flat field spectrum'' was then heavily smoothed and was
afterwards used to normalize the dome flats along the dispersion axis. 

For the wavelength calibration we fitted 3$^{\rm rd}$ order polynomials to
the dispersion relations. We rebinned the frames two-dimensionally to
constant wavelength steps. Before the sky fit the frames were median
filtered along the spatial axis to erase cosmic ray hits in the 
background. To
determine the sky background we had to find regions without any stellar
spectra, which were sometimes not close to the place of the object's
spectrum. Nevertheless the flat field correction and wavelength calibration
turned out to be good enough that a constant spatial distribution fit the
sky light well enough to subtract the sky background at the object's
position with sufficient accuracy. This means in our case that we do not
see any absorption lines caused by the predominantly red stars of the
clusters or by moon light. The fitted sky background was then subtracted
from the unsmoothed frame and the spectra were extracted using Horne's
algorithm (Horne \cite{horn86}) as implemented in MIDAS. 

Finally the spectra were corrected for atmospheric extinction using the
data of T\"ug (\cite{tueg77}). The data for the flux standard stars were
taken from Hamuy et al. (\cite{hamu92}). The response curves were fitted by
8$^{\rm th}$ order polynomials for the low and medium resolution spectra.
We took special care to correctly fit the response curves for the low
resolution data in the region of the Balmer jump, since we planned to use
this feature for the determination of the effective temperatures of the
stars. The response curves showed a roughly constant shape, but varied in
intensity, indicating non-photometric conditions during the observations. 

We also used the medium resolution data to derive radial velocities, which
are listed in Table~\ref{47tuc_n362_targ} (corrected to heliocentric
system). The error of the velocities is about 30~km/sec (estimated from the
r.m.s. scatter of the velocities derived from individual lines). The
normalized and velocity-corrected spectra are plotted in
Fig.~\ref{47tuc_n362_spectra}. 

\section{Atmospheric Parameters\label{47tuc_sec_par}}

\subsection{\teff\ derived from energy distributions\label{47tuc_sec_energy}}

To avoid any spurious results due to contamination from neighbouring
stars we searched all stars for red neighbours. For this purpose we used
the photometry of 47~Tuc and NGC~362 (Montgomery \& Janes, priv. comm.),
and extracted for each of our targets all stars within a radius of
20\arcsec\ (down to the limiting magnitude of $B\approx$~\magpt{18}{5}). We
then estimated the straylight provided by these stars assuming that the
seeing at larger distances is best described by a Lorentz profile and
scaling the intensities with the V fluxes. We used 1\bsec5 as ``intrinsic
seeing'' for all observations (which overestimates the seeing near the
meridian) and took into account the elongation caused by atmospheric
dispersion. Under these assumptions we get stray light levels of more than
3\% in $B$ only for {\it MJ3832} (36\%) in NGC~362. 

We used line blanketed LTE model atmospheres of Kurucz (\cite{kuru92},
ATLAS9) for a metallicity [M/H] = $-$1.0 to derive effective temperatures
from the low resolution spectra. To correct for interstellar reddening we
applied the reddening law of Savage \& Mathis (\cite{sama79}) and used an
E$_{\rm B-V}$ of \magpt{0}{04} for both clusters. 

The simultaneous fitting of the spectrophotometric data (Balmer jump and
continuum) was generally possible except for {\it MJ38529} in 47~Tuc and
{\it MJ6558} and {\it MJ3832} in NGC~362. {\it MJ3832} has a cool star
close by ({\it MJ3749}, distance 3\arcsec, $B$ = \magpt{14}{39}, $B-V$ =
\magpt{+1}{43}) which contaminates the low resolution spectrum that was
observed with a 5\arcsec\ slit (see above). {\it MJ6558} was observed at
rather high airmass (1.8) and {\it MJ38529} is probably a binary (see
below). However, comparing our measured intensities to those derived from
$B$ and $V$ magnitudes of the stars led to discrepancies of up to
\magpt{0}{3}. In most of the cases the observed spectra were brighter than
the corresponding photometric fluxes. The slope as derived from $B-V$ was
consistent with the slope of the spectrophotometric continuum in most cases
except {\it MJ38529} in 47 Tuc and {\it MJ5381} in NGC~362 ($B-V$ being
more positive than indicated by the spectrophotometric continuum). These
variations are consistent with non-photometric weather conditions, assuming
that the varying atmospheric absorption was basically grey, i.e.
independent of wavelength. If the absorption towards the flux standard
stars were greater than towards the globular clusters, the cluster stars
would be calibrated too bright. We found that the observed H$_\beta$ line
in the low resolution spectrum of {\it MJ8453} in NGC~362 was significantly
shallower than the theoretical one (see also below). 

\subsection{Balmer line profile fits\label{47tuc_sec_balmer}} 

To derive effective temperatures, surface gravities, and helium abundances
we fitted the observed Balmer and helium lines with appropriate stellar
model atmospheres. Beforehand we corrected the spectra for radial velocity
shifts, derived from the positions of the Balmer lines. To establish the
best fit we used the routines developed by Bergeron et al. (\cite{besa92})
and Saffer et al. (\cite{saff94}), which employ a $\chi^2$ test. The
$\sigma$ necessary for the calculation of $\chi^2$ is estimated from the
noise in the continuum regions of the spectra. The fit program normalizes
model spectra {\em and} observed spectra using the same points for the
continuum definition. We computed model atmospheres using ATLAS9 (Kurucz
1991, priv. comm.) and used Lemke's version\footnote{For a description see
http://a400.sternwarte.uni-erlangen.de/$\sim$ai26/linfit/linfor.html} of
the LINFOR program (developed originally by Holweger, Steffen, and
Steenbock at Kiel university) to compute a grid of theoretical spectra,
which include the Balmer lines H$_\alpha$ to H$_{22}$ and \ion{He}{i}
lines. The grid covered the ranges 8000~K$\le$ \teff\ $\le$20,000~K,
2.5$\le$ \logg\ $\le$6.0, and $-$2.0, $-$1.0 in \loghe\ at metallicities of
$-$1 and $-$0.75. 

We fitted the Balmer lines from H$_\beta$ to H$_{12}$ (excluding
H$_\epsilon$ because of the \ion{Ca}{ii} H line) and the \ion{He}{i} lines
$\lambda\lambda$ 4026~\AA , 4388~\AA , 4472~\AA, 4713~\AA , and 4922~\AA\
(for \teff\ $\ge$ 10,500~K). For cooler stars we fitted only the Balmer
lines for a fixed helium abundance of \loghe\ = $-$1 (the helium abundance
in these cooler stars should be close to solar as they should not be
affected by diffusion). In the medium resolution spectrum of {\it MJ8453}
the observed H$_\beta$ line is significantly less deep than the theoretical
one and also H$_\gamma$ shows some evidence for filling. We therefore
excluded those two lines from the fit for this star. Several stars above
10,500~K do not show \ion{He}{i} lines. We nevertheless fitted those
regions of the spectra, where the \ion{He}{i} lines are expected, to obtain
an upper limit of the He abundance. 

To check for any effects of metallicity on the final results we fitted the
spectra with models of [M/H] = $-$1.00 and [M/H] = $-$0.75. We found that
in all cases except {\it MJ3832} in NGC~362 the differences in effective
temperature and surface gravity were below 1\% resp. 0.05~dex (2\% and
0.11~dex for {\it MJ3832}). As the differences are below our expected
errors we decided to use only the results for [M/H] = $-$1 for the further
discussion, which lies between the metallicity for 47~Tuc ([M/H] = $-$0.71) 
and that of NGC~362 ([M/H] = $-$1.16). 

The fit program gives r.m.s. errors derived from $\Delta \chi^2$ = 2.71
(\teff , \logg) resp. 3.53 (\teff , \logg , \loghe). However, if
$\chi^2$ is not close to 1 (see Table~\ref{47tuc_n362_par}) these errors
will most likely underestimate the true errors. As the errors causing
$\chi^2>1$ are most probably systematic (see below) they are hard to
quantify. We decided to get an estimate of their size by assuming that the
large $\chi^2$ is solely caused by noise. Increasing the noise parameter
$\sigma$ until $\chi^2$ = 1 then yields new formal errors for $\Delta
\chi^2$ = 2.71 and 3.53, respectively, which are given in
Table~\ref{47tuc_n362_par}. These errors still underestimate the internal
errors as we oversampled the spectra by a factor of 2.5 when rebinning to
constant wavelength steps. Thus only 40\% of the wavelength points used for
the fit are truly independent. We correct this by applying a factor of
$\sqrt{2.5} = 1.6$ to all error estimates. We believe that the $\chi^2$
values above 1 are due to the low resolution of the data (5.7~\AA): At
\teff\ = 11,000~K and \logg\ = 4 the FWHM of the theoretical Balmer lines
are around 12~\AA . Thus the instrumental profile makes up a considerable
part of the observed line profile. Therefore any deviations of the
instrumental profile from a Gaussian (which is used to convolve the model
spectra) will lead to a bad fit. Fortunately we could compare the effects
for one hot HB star (B~3253) in NGC~6752, for which we have an NTT spectrum
with the same setup as is used here and a spectrum from the ESO 1.52m
telescope with a resolution of 2.6~\AA. While the reduced $\chi^2$ values
for the two spectra differ by a factor of 2 (5.9 vs. 3.1) the resulting
effective temperatures and surface gravities are rather similar
(13,700~K/3.75 vs. 13,700~K/3.80, see Moehler et al. \cite{mosw00} for
details). 

\begin{table*}
\caption[] {Atmospheric parameters and masses for the programme stars as
derived from low and medium resolution spectroscopic data. The surface
gravities derived from the low resolution spectrophotometric data are
rather uncertain. We also give the reduced $\chi^2$ values from the line
profile fits and the errors listed below are the r.m.s. errors of the fit
routine adjusted as described in the text. The three rightmost columns
give the cluster resp. SMC membership according to the radial velocity,
proper motion, and derived mass of the star (see Sect.~\ref{47tuc_sec_disc}
for details). A --- means that the information places a star neither to the
globular cluster nor to the SMC. Brackets note dubious assignments.
\label{47tuc_n362_par}}
\begin{tabular}{l|rr|rrrl|rr|ccc}
\hline
Star & \multicolumn{2}{c|}{spectrophotometry} &
\multicolumn{4}{c|}{medium resolution data} &
\multicolumn{2}{c}{Masses} & \multicolumn{3}{|c}{membership}\\
    & \teff &        \logg  & $\chi^2$ & \teff  & \logg  & \loghe & 
 cluster & SMC & v$_{\rm rad}$ & proper & mass\\
    & [K] & [cm/s$^2$] & & [K] & [cm/s$^2$] & & [\Msolar] & [\Msolar] 
 & & motion & \\
\hline
\multicolumn{12}{c}{ 47Tuc  }\\
\hline
 {\it MJ280}   & 13500 & 4.0: & 2.75 & 14500$\pm$290 & 4.21$\pm$0.08 & 
$-$1.41$\pm$0.16 & 0.65 & 133 & C & C & C \\
 {\it MJ8279}  & 18000 & 4.5: & 1.96 & 18500$\pm$530 & 4.20$\pm$0.10 & 
 $-$1.56$\pm$0.11 & 0.02 & 3.8 & SMC &  & SMC\\
 {\it MJ33410} & 10000 & 3.5: & 4.53 & 10400$\pm$210 & 3.53$\pm$0.11 & 
$-$1.00 & 0.51 & 104 & C &  & C \\
 {\it MJ38529} &  8000$^1$ & 3.0: & 2.25 &  7950$\pm$20 & 5.21$\pm$0.05 & 
 $-$1.00 & 24 & 4900 & C & C & ---\\
         & 12500$^2$ & 3.5: & 7.85 & 14000$\pm$400 & 5.34$\pm$0.11 & 
$<$$-$2 & 8.4 & 1700 & C & C & ---\\
\hline
\multicolumn{12}{c}{ NGC362 }\\
\hline
 {\it MJ65}    &  8500 & 2.5: & 2.03 &  9460$\pm$540 & 2.46$\pm$0.32 & 
$-$1.00 & 0.16 & 9.5 & SMC & --- & SMC\\
 {\it MJ94}    & 11000 & 3.5: & 3.53 & 11700$\pm$350 & 3.33$\pm$0.11 & 
$-$1.93$\pm$0.35 & 0.13 & 7.6 & SMC & --- & SMC\\
 {\it MJ2341}  & 16000 & 4.5: & 2.08 & 17400$\pm$500 & 3.97$\pm$0.10 & 
$-$1.79$\pm$0.13 & 0.10 & 5.9 & --- &  & SMC\\
 {\it MJ3832}  &       &      & 2.64 &  8250$\pm$140 & 2.77$\pm$0.06 & 
$-$1.00  & 0.23 & 13 & C & & (C)\\
 {\it MJ5381}  &  9000 & 3.0: & 2.43 &  7780$\pm$100 & 2.14$\pm$0.05 & 
$-$1.00 & 0.03 & 2.0 & (SMC) & --- & ---\\
 {\it MJ6558}  &       &      & 3.06 & 12500$\pm$350 & 4.09$\pm$0.11 & 
$-$1.61$\pm$0.40  & 1.08 & 62 & C & --- & (C) \\
 {\it MJ8241}  &  8500 & 3.5: & 3.49 &  7980$\pm$50 & 3.06$\pm$0.06 & $-$1.00 
 & 0.51 & 29 & C & & C\\
 {\it MJ8453}$^3$  & 14000 & 4.0: & 1.53 & 16600$\pm$560 & 4.11$\pm$0.11 & 
$-$1.99$\pm$0.11 & 0.27 & 18 & --- & --- & (C) \\
\hline
\end{tabular}
\begin{tabular}{l}
$^1$ fitting the continuum\\
$^2$ fitting the Balmer jump\\
$^3$ H$_\beta$, H$_\gamma$ are not included in the fit of {\it MJ8453}\\
\end{tabular}
\end{table*}

\begin{figure}
\vspace{7.5cm}
\includegraphics{10063.f3}
\caption[]{The atmospheric parameters of the target stars in 47~Tuc and
NGC~362 compared to zero-age horizontal branch models (Dorman et al.
\cite{dorm93}, solid and dashed lines) and (post)-main-sequence models
(Charbonnel et al. \cite{chme93}, dotted lines). The numbers along the
main sequence give the mass of the model star in solar masses. The
suspected field/SMC stars are marked by open symbols.
\label{47tuc_n362_tg}} 
\end{figure}

\subsection{Comparison of effective temperatures obtained with different 
methods\label{47tuc_sec_compare}}

As can be seen from Table~\ref{47tuc_n362_par} the temperatures determined
 from the spectrophotometric data deviate from those derived
from the Balmer lines, which may be an effect of strongly varying
atmospheric extinction (as is already suggested by the discrepancies
between spectrophotometric and photometric fluxes for the stars, see
Sect.~\ref{47tuc_sec_energy}). Differences in the reddening and
interstellar extinction law between galactic globular clusters and the SMC
may also affect the flux distribution of the stars, which are by default
only corrected for the galactic extinction towards the respective cluster.
Sasselov et al. (\cite{sabe97}) report that the extinction law for the SMC
is essentially the same as that for the Large Magellanic Cloud (LMC). We
therefore dereddened the extinction corrected spectra once more, this time
using \ebv\ = \magpt{0}{09} and the reddening law towards the LMC (Howarth
\cite{howa83}) as implemented in MIDAS. The resulting differences between
spectra with and without this additional reddening correction are too small
to explain the temperature differences between the results from the low
resolution data and the line profiles. We thus conclude that the most
probable explanation for the discrepancies lies with the non-photometric
observing conditions. 

\subsection{Masses\label{47tuc_ssec_mass}} 

Knowing \teff , \logg , and the distances of the stars we can derive the
masses: 

\noindent $\log{\frac{\rm M}{\rm M_\odot}} = const.+\log
g+0.4\cdot((m-M)_V-V+V_{th})$ 

\noindent (V$_{th}$ denotes the theoretical brightness at the stellar
surface as given by Kurucz \cite{kuru92}.) The results are listed in
Table~\ref{47tuc_n362_par}. We determine two masses for each star, one
assuming that it is a cluster member and one assuming it belongs to the
SMC. We use the following reddening-free distance moduli $(m-M)_0$:
\magpt{13}{31} for 47~Tuc, \magpt{14}{67} for NGC~362 (both from Djorgovski
\cite{djor93}; Harris \cite{harr96} gives \magpt{13}{25} resp.
\magpt{14}{65} in the June 22, 1999 version of his compilation), and
\magpt{18}{91} for the SMC (Sasselov et al. \cite{sabe97}). Using the {\sc
Hipparcos} based distance moduli to 47~Tuc and NGC~362 of \magpt{13}{51}
and \magpt{14}{88} (Gratton et al. \cite{grfu97}, Reid \cite{reid98}) would
increase the masses by about 20\% in both cases. 

In addition to the error in \logg\ given in Table~\ref{47tuc_n362_par}
errors in the absolute magnitude and the theoretical brightness at the
stellar surface also enter the final error in \logM . We assume an uncertainty
 in the absolute brightness (combining errors in the
photometric data, uncertainties in the distance moduli and
reddenings) of \magpt{0}{3}, and a mean error in \logg\ of 0.2~dex. 
The error in the theoretical $V$ brightness is dominated by the
errors in \teff . We estimate the total error in \logM\ to be about
0.24~dex, corresponding to an error of $^{+73\%}_{-42\%}$ 
in mass (for a
detailed discussion of error sources see Moehler et al. \cite{mohe97}). 

\section{Discussion\label{47tuc_sec_disc}} 

{\bf 47~Tuc:\label{47tuc_disc}}
\ As can be seen from Table~\ref{47tuc_n362_par} three of the four stars in
47~Tuc are probably members of the cluster, but {\it MJ8279} according to its
radial velocity belongs to the SMC. The comparison of the spectroscopically
derived mass (3.8\Msolar) with the mass obtained from the evolutionary
tracks (5\Msolar) further supports its SMC membership.  Kaluzny et al.
(\cite{kakr97},\cite{kawy98}) reported observations of six stars in the field of
47~Tuc with similar
magnitudes and colors to {\it MJ 8279}, that is $18 <$ V $< 18.7$, and $B-V 
\approx
-0.1$.    The evidence of {\it MJ 8279} supports their conclusion (based on a
lack of a radial gradient) that most of these stars are probably SMC members.

Of the remaining
three stars {\it MJ38529} is very probably a binary: Its UV-visual colour
suggests an effective temperature of 11,900~K, but $B-V$ for this star is
much too red\footnote{Kaluzny et al. (\cite{kaku98}) give a relatively red 
$V-I$ colour of $\approx +0.14$ for {\it MJ38529}, which provides further 
evidence for the presence of a cool companion.} for such
a temperature (see also its position in Fig.~\ref{cmd_47tuc_n362} below the
HB). The optical low resolution spectrum suggests effective
temperatures of 12,500~K (Balmer jump only) resp. 8000~K (continuum only).
The superposition of two different stellar spectra (with different Balmer
lines) may explain the strange values obtained for \logg\ from the Balmer
line profiles. If we assume that the temperature obtained from UV data
is that of the hot component {\it MJ38529hot}, it would lie between 
{\it MJ280} and {\it
MJ33410} in \teff. If {\it MJ38529hot} were a ZAHB star, its $V$ magnitude
should also lie between those two stars -- in contradiction to the fact that
the complete binary system is already fainter in $V$ than {\it MJ280} and
{\it MJ33410}. {\it MJ38529hot} could thus also be an extreme blue straggler,
which would be fainter than a ZAHB star of the same \teff . On the other
hand, the UV-based temperature is a lower limit for the hot component, as
optical data are also involved in its determination (UV-visual colour). If
{\it MJ38529hot} were considerably hotter than {\it
MJ280} it would be much fainter visually and could therefore be a hot ZAHB 
star accompanied by a red subgiant.
Thus the contamination by the cool star prevents any definite conclusions. 

{\it MJ280} is hot enough to be affected by diffusion processes which may
result in an enrichment of heavy elements in its atmosphere (cf. Moehler et
al. \cite{mosw00}). Fitting it with enriched model atmospheres as described by
Moehler et al. (\cite{mosw00}, [Fe/H] = $+$0.5) indeed yields \teff\ =
14,200~K, \logg\ = 4.29 and \loghe\ = $-$1.63, moving the star closer to
the ZAHB. 

{\it MJ33410} finally is a normal cool HBB star. The masses of
{\it MJ280} and {\it MJ33410} agree well with canonical predictions. 

{\bf NGC~362:\label{ngc362_disc}}
\ According to Table~\ref{47tuc_n362_targ} three of the eight targets in
NGC~362 are radial velocity members. {\it MJ6558} is not a proper motion
member, but the error of its proper motion is rather large compared to the
measurements of the other stars. {\it MJ6558} and {\it MJ8241} lie close to
the canonical ZAHB, whereas {\it MJ3832} has evolved towards lower \logg.
The mean mass of all three stars is 0.50$^{+0.44}_{-0.23}$\Msolar\ (r.m.s. 
errors only) and thus also in good agreement with canonical HB theory. 
Note that MJ6558 was one of the three candidate blue HB stars (= H1382) 
in NGC~362 identified by Harris et al. (\cite{harr82}); the other two Harris
candidates were not detected with UIT and radial velocity studies indicate that
they are unlikely to be cluster members (R. Rood, priv. comm.).  
None of the radial velocity members are good candidates for being extreme blue
stragglers, in the sense of having both a mass and  \teff\ consistent with being
on an extension of the NGC~362 main sequence.

The three radial velocity member stars are all located within 2\bmin5 
of
the cluster center, while the remaining five stars are all more than 3\bmin5
from the  center.
Three of these five stars may belong to the SMC according to their radial 
velocities:
{\it MJ65}, {\it MJ94}, and {\it MJ5381}. They lie, however, farther from
the SMC main sequence in Fig.~\ref{cmd_47tuc_n362} than the remaining
two stars {\it MJ2341} and {\it MJ8453}. Comparing the spectroscopically
determined masses to those derived from evolutionary sequences would put
{\it MJ65} (9.5\Msolar\ vs. 7\Msolar), {\it MJ94} (7.6\Msolar\ vs.
4\Msolar), and {\it MJ2341} (5.9\Msolar\ vs. 5\Msolar) to the SMC.  
The radial velocity of {\it MJ2341} differs by 74 km~s$^{-1}$ from that of 
the SMC, and so it might be a high-velocity halo star of the SMC
(see, e.g., Keenan \cite{keen92}, \cite{keen97}
for a discussion of main-sequence B stars in the halos of galaxies).  While
two of the three stars lie away from the main sequence one should keep
in mind that our selection is heavily biased towards bright (= evolved) hot
SMC stars.   Note that {\it MJ65} and {\it MJ94} appear to be normal HB 
stars in
the UV CMD of Dorman et al. (\cite{dosh97}), which emphasizes the need for
spectroscopic observations to verify the evolutionary status of stars in fields
with high contamination.

The evolutionary status of {\it MJ5381} is unclear: In the $m_{\rm
162}, m_{\rm 162}-V$ diagram of Dorman et al. (\cite{dosh97}) it lies
rather isolated -- being fainter than the theoretical ZAHB for NGC~362, but
much cooler than the SMC main sequence. 

The filled H$_\beta$ line of {\it
MJ8453} reminds us of the Be stars analysed by Mazzali et al.
(\cite{male96}), but it is considerably cooler than those objects. A
spectrum of the H$_\alpha$ region of this star would be helpful to
decide whether it indeed shows a Be-like spectrum. 

\section{Conclusions\label{47tuc_ssec_conc}}

There has been increasing interest in understanding the origin of the hot
stars that can appear in old, metal-rich systems. The metal-rich open
clusters NGC~6791 ([Fe/H] = +0.4) and NGC~188 ([Fe/H] = 0.0) contain hot HB
populations sufficiently large that the UV-turnup in their integrated
cluster spectra would be as strong as that observed in elliptical
galaxies (Landsman et al. \cite{labo98}). At these high metallicities, hot HB
stars can be a natural product of single-star HB models (e.g. Dorman et al.
\cite{dorm95}), although a significant fraction
of the hot stars in the two clusters are binaries (Green et al.
\cite{grli97}). The metal-rich bulge globular clusters, NGC~6388 
([Fe/H] = $-$0.60, Harris \cite{harr96}) and NGC~6441 ([Fe/H = $-$0.53, 
Harris \cite{harr96}) contain a sizable population of blue HB
stars. A possible scenario for the blue HB in these clusters is a merger of
two systems with different ages, with the blue HB arising from the minority
old population (Yoon et al. \cite{yole00}). However, this scenario does not
explain several other peculiar aspects of the blue HB population in
NGC~6388 and NGC~6441, including the slope of the HB, and the gravities
which are higher than predicted by canonical models (Moehler et al.
\cite{mosw99}). 

Here we have shown that at least some hot HB stars exist in the classical red
HB clusters 47~Tuc and NGC~362, and -- unlike the case of NGC~6388 and NGC~6441
 -- they perfectly fit the expectations of
canonical HB evolution.     The three blue members in 47 Tuc all have 10,000 K
$<$ \teff $<$ 15,000 K, and thus are well separated from the rest of the HB
population, which is (except for the single RR Lyr V9) entirely redward of the
instability strip.    (A fourth probable hot star member of 47 Tuc is UIT-14,
which is only $1.7'$ from the cluster center, and  for which the IUE spectrum
obtained by  O'Connell et al. \cite{ocdo97} indicates \teff $\approx$
50,000~K.) The small number of hot HB stars in 47~Tuc, and their high
temperatures, point to a scenario in which they have a different
physical origin (such as binary interactions) than the dominant red HB
population. It is suggestive that one of the hot stars in 47 Tuc
({\it MJ38529}) appears to be a photometric binary, whose hot 
component could, however, also be an extreme blue straggler.

Interestingly, whereas NGC~362 has 14 bright blue stars within 14\arcsec\ radius
of its center (Dorman et al. \cite{dosh97}), the inner 1\arcmin\ of 47 Tuc
probably does {\em not} contain any hot HB stars,  as shown by observations
with IUE (Rich et al. \cite{rimi93}), UIT (O'Connell et al. \cite{ocdo97}), and
WFPC2 (Rich et al. \cite{riso97}).   Rose \& Deng (\cite{rode99}) found that
only about 7\% of the mid-UV light in the core of 47~Tuc comes from stars
hotter than about 7,500~K (most of which are probably blue stragglers). 

The three hot HB stars in NGC~362 are somewhat cooler than those in 47~Tuc, and
the dominant HB in NGC~362 is
not as red as that of 47~Tuc.   Thus it is plausible that the blue HB stars in
NGC~362 arise from a small percentage of red giants with unusually high mass
loss, and no special mechanism is required for their production.  
Unfortunately, the SMC contamination is larger in NGC~362 than in 47~Tuc, and
five out of our eight targets turned out to be nonmembers.  It would be
interesting to study the stellar parameters of the hot stars in the
core region, which we could not observe due to high crowding, but where the
relative SMC contamination should be much lower.

\acknowledgements
We want to thank the staff of the ESO La Silla observatory for their
support during our observations and an anonymous referee for valuable 
comments. SM acknowledges support by the DLR (grant
50~OR~96029-ZA).

\end{document}